\documentclass[journal,twoside,web]{ieeecolor}
\usepackage{generic}
\usepackage{cite}
\usepackage{amsmath,amssymb,amsfonts}
\usepackage{algorithmic}
\usepackage{graphicx}
\usepackage{algorithm,algorithmic}
\usepackage{hyperref}
\hypersetup{hidelinks=true}
\usepackage{textcomp}
\usepackage{booktabs}
\usepackage{svg}
\usepackage[table]{xcolor}
\usepackage{pdfpages}
\usepackage{amsmath}
\usepackage{ragged2e}
\usepackage{comment}
\usepackage[acronym]{glossaries}
\usepackage{acronym}
\usepackage{subcaption}
\def\BibTeX{{\rm B\kern-.05em{\sc i\kern-.025em b}\kern-.08em
    T\kern-.1667em\lower.7ex\hbox{E}\kern-.125emX}}

\begin{document}
\acrodef{EMG}{Electromyography}
\acrodef{ECG}{electrocardiography}
\acrodef{PPG}{photoplethysmography}
\acrodef{EEG}{electroencephalography}
\acrodef{EOG}{electrooculography}
\acrodef{SoA}{State-of-the-Art}

\acrodef{FM}{Foundation Models}

\acrodef{MCU}{microcontroller unit }

\acrodef{RMSE}{Root Mean Squared Error}
\acrodef{R}{Pearson Correlation Coefficient}
\acrodef{PSD}{Power Spectral Density}
\title{TinyMyo: a Tiny Foundation Model for Flexible EMG Signal Processing at the Edge}
\author{Matteo Fasulo, Giusy Spacone, \IEEEmembership{Graduate Student Member, IEEE}, Thorir Mar Ingolfsson, \IEEEmembership{Member, IEEE}, Yawei Li, \IEEEmembership{Member, IEEE}, Luca Benini, \IEEEmembership{Fellow, IEEE}, and Andrea Cossettini, \IEEEmembership{Senior Member, IEEE}
\thanks{This work was supported in part by the ETH Future Computing Laboratory (EFCL), by a donation from Huawei Technologies, and by a grant from the Swiss National Supercomputing Centre (CSCS) under project ID lp12 on Alps. We also acknowledge ISCRA for awarding this project access to the LEONARDO supercomputer, owned by the EuroHPC Joint Undertaking, hosted by CINECA (Italy).}
\thanks{M. Fasulo, G. Spacone, T. M. Ingolfsson, Y. Li, A. Cossettini, and L. Benini are with the Integrated Systems Laboratory of ETH Z{\"u}rich, Z{\"u}rich, Switzerland (\texttt{gspacone@iis.ee.ethz.ch}).}
\thanks{L. Benini is also with the Department of Electrical, Electronic and Information Engineering (DEI), University of Bologna, Bologna, Italy.}
}
\makeatother
\makeatletter
\def\ps@mynotice{%
  \def\@oddfoot{%
    \hfil
    \parbox[t]{\textwidth}{\scriptsize\vspace{-0.5cm}
      \copyright\ 2026 IEEE.  Personal use of this material is permitted.  Permission from IEEE must be obtained for all other uses, in any current or future media, including reprinting/republishing this material for advertising or promotional purposes, creating new collective works, for resale or redistribution to servers or lists, or reuse of any copyrighted component of this work in other works.%
    }%
    \hfil
  }%
  \def\@evenfoot{\@oddfoot}%
}
\makeatother

\maketitle
\thispagestyle{mynotice}   


\makeatletter
\let\old@bibitem\@bibitem
\renewcommand{\@bibitem}[1]{%
  \ifboolexpr{
    test {\ifstrequal{#1}{salter_emg2pose_2024}} or
    test {\ifstrequal{#1}{chen_2025_physiowave}} or
    test {\ifstrequal{#1}{goswami_2024_moment}} or
    test {\ifstrequal{#1}{doner2025luna}} or
    test {\ifstrequal{#1}{su_2023_roformerenhancedtransformerrotary}} or
     test {\ifstrequal{#1}{jiang_2024_labram}} or
    test {\ifstrequal{#1}{pmlr-v139-touvron21a}} or
    test {\ifstrequal{#1}{tegon_2025_femba}} or
    test {\ifstrequal{#1}{gaddy-klein-2021-improved}} or
    test {\ifstrequal{#1}{baevski_2020_wav2vec}}
  }{\color{black}}{\color{black}}
  \old@bibitem{#1}%
}

\let\old@lbibitem\@lbibitem
\renewcommand{\@lbibitem}[2]{%
  \ifboolexpr{
    test {\ifstrequal{#2}{salter_emg2pose_2024}} or
    test {\ifstrequal{#2}{chen_2025_physiowave}} or
    test {\ifstrequal{#2}{goswami_2024_moment}} or
    test {\ifstrequal{#2}{doner2025luna}} or
    test {\ifstrequal{#2}{su_2023_roformerenhancedtransformerrotary}} or
    test {\ifstrequal{#2}{jiang_2024_labram}} or
    test {\ifstrequal{#2}{pmlr-v139-touvron21a}} or
    test {\ifstrequal{#2}{tegon_2025_femba}} or
    test {\ifstrequal{#2}{gaddy-klein-2021-improved}} or
    test {\ifstrequal{#2}{baevski_2020_wav2vec}}
  }{\color{black}}{\color{black}}
  \old@lbibitem[#1]{#2}%
}
\makeatother

\begin{abstract} 
Objective: Surface electromyography (EMG) is a non-invasive sensing modality used in biomechanics, rehabilitation, prosthetic control, and human-machine interfaces. However, achieving robust generalization across subjects, recording systems, and acquisition protocols remains challenging. While foundation models (FMs) are gaining traction for EMG, existing approaches remain limited to single downstream tasks and lack deployability. We address these limitations.
Methods: We present TinyMyo, a lightweight FM based on a Transformer encoder architecture. The model is pre-trained in a self-supervised manner using masked reconstruction on public datasets.
With 3.6M parameters, TinyMyo is designed to support multiple downstream tasks through minimal task-specific head adaptations.
Results:
We demonstrate generalization across hand gesture classification, hand kinematic regression, speech production and recognition, with performance comparable or surpassing the state of the art (SoA), and model size below 5M parameters. We achieve SoA results compared to previous FM-based works on the \textit{NinaPro DB5} (87.98\%), \textit{UCI-EMG} (97.1\%), and \textit{EPN-612} (96.6\%) datasets. 
We demonstrate the first-time deployment of an EMG FM on an ultra-low power microcontroller (GAP9), with an inference time of 0.785 s, energy of 44.91 mJ and power envelope of 57.18 mW, reduced to 0.496 s and 0.089 s inference times with a smaller (1.9M) model variant on 5 and 1-s data windows.
Conclusion: TinyMyo demonstrates that compact, self-supervised EMG FM can guarantee generalization across multiple downstream tasks while remaining compatible with low-power edge devices. 
Significance: 
TinyMyo is the first EMG FM for ultra-low-power edge-devices, enabling scalable and energy-efficient sensing for motor intent decoding, neuromuscular assessment, and human–machine interaction.
\end{abstract}

\begin{IEEEkeywords}
EMG, self-supervised, foundation model, edgeAI, gesture recognition, kinematic regression, speech
\end{IEEEkeywords}
\section{Introduction}\label{sec:intro}
\bstctlcite{IEEEexample:BSTcontrol}
\ac{EMG} measures variations in the electrical activity of muscles during contractions, reflecting the motor-neuron activation patterns~\cite{farina_2023_rev}. \ac{EMG} can be acquired using invasive methods (needle \ac{EMG}), suitable for a detailed characterization of individual motor units, or using non-invasive, surface-based techniques, to capture the global activity of muscle fibres ~\cite{farina_2023_rev, avila_2023_rev}.

Surface \ac{EMG} has found extensive application across a wide range of domains, including biomechanical analysis and rehabilitation~\cite{avila_2023_rev}, human–robot interaction~\cite{xiong_emgrobot_2024}, prosthetic control
\cite{ marinelli_2023emgupper}, and emerging human–computer interaction interfaces~\cite{generic_neuromotor_interface_2025}.
However, despite decades of widespread use, many important challenges remain. 

\ac{EMG} signals are strongly affected by noise, motion artifacts, cross-talk, and acquisition variability across electrode layouts, hardware platforms, and acquisition pipelines \cite{boyer_emg_noise,merletti_2020_tutorial}.
In addition, both inter-subject and intra-subject factors \cite{xiong_2021_review}, including anatomy, electrode placement, skin impedance, and physiological variability lead to substantial distribution shifts, hindering generalization~\cite{ni_2024_survey}. Most of these noise and variability sources can only be partially compensated at the signal acquisition level and instead require dedicated algorithmic approaches. However, designing models that generalize across subjects, devices, and acquisition conditions remains a significant challenge.

\ac{FM} offer a great promise to address these challenges. Inspired by the success of self-supervised pre-training in vision, language, and speech, as well as by emerging applications in biosignals \cite{gu_2025_foundationsurvery}, the core hypothesis is that large-scale datasets can be leveraged to learn high-level, platform and subject-agnostic signal representations. For \ac{EMG}, such models could provide a unified backbone for many downstream tasks, building upon a robust, device-agnostic feature extraction, improving generalization across devices and subjects. In this context,  the key idea is that large-scale pre-training can enable the learning of invariant representations that mitigate the impact of noise, interference, and variability.
However, progress in \acp{FM} of \ac{EMG} has been limited by the scarcity of large corpora, and only with recent large-scale datasets (collected from thousands of users) \cite{salter_emg2pose_2024}, the exploration of generalized \acp{FM} of \ac{EMG} became viable.

In this context, existing \ac{EMG} \acp{FM} present significant limitations. Current models mostly focus on a single task (typically gesture classification) \cite{chen_2025_physiowave,chen_2025_waveformer}. Furthermore, the size of \ac{EMG} \acp{FM}  remains large, and no prior work demonstrated the deployment on embedded hardware. However, for many real-life applications, such models must operate under the strict resource constraints of wearable devices. Effective deployment in real-world scenarios requires not only accuracy and robustness, but also compactness and computational efficiency to meet stringent memory, power, and latency constraints.

We address these limitations via the following contribution.
\begin{itemize}
    \item We present \textbf{TinyMyo}, a lightweight Transformer-based encoder operating directly on \ac{EMG} signals in the time domain. TinyMyo is pre-trained using a self-supervised masked-reconstruction framework on publicly available, heterogeneous \ac{EMG} datasets, achieving high reconstruction fidelity with only \textbf{3.6 million} parameters.
    \item Unlike previous \ac{EMG} \acp{FM} , which are typically designed around a single downstream task or rely on large, domain-agnostic encoders, TinyMyo \textbf{generalizes across heterogeneous \ac{EMG} datasets and sensor configurations}. Our architecture follows the general Transformer design adopted in prior \ac{EMG} works~\cite{chen_2025_physiowave, chen_2025_waveformer}, formulated in a \emph{unified \ac{EMG}} \ac{FM} able to support a wide range of downstream tasks using the same pretrained backbone. We demonstrate the versatility of the pre-trained encoder through minimal task-specific head adaptations across diverse downstream tasks, including \textbf{hand-gesture classification} on the Ninapro DB5 \cite{ninapro_db5}, EPN-612 \cite{benalcazar_emg-epn-612_2020}, UC Irvine \cite{emg_data_for_gestures_481} and Meta's Generic Neuromotor Interface \cite{generic_neuromotor_interface_2025} datasets, \textbf{hand-kinematic regression} on the Ninapro DB8 \cite{ninapro_db8}, and \textbf{speech} production and discrimination on the Gaddy Silent Speech Dataset \cite{gaddy-klein-2020-digital}. Our model achieves competitive performances across all downstream tasks, either surpassing \ac{SoA} or guaranteeing a comparable accuracy with a sub-5M class model size.
    \item We demonstrate for the first time the \textbf{deployment of an \ac{EMG} \ac{FM} on an ultra-low-power microcontroller} (GAP9), achieving an inference time of $0.785~\text{s}$, energy of $44.91~\text{mJ}$ and power envelope of $57.18~\text{mW}$. A smaller model variant (TinyissimoMyo, 1.9M parameters) reduces inference time to 0.496 s and 0.089 s on 5-s and 1-s windows, respectively, reaching the sub-100 ms latency regime required by real-time applications such as prosthetic control, at the cost of reduced accuracy on complex classification benchmarks.
    \item We open source the model and the weights to enable reproducibility and further research by the community (\url{https://github.com/pulp-bio/BioFoundation}).
\end{itemize}

\section{Related Works}\label{sec:related}

Several efforts have been made to develop \acp{FM} for biosignals, and most of the works focused on \ac{ECG}, \ac{PPG} and \ac{EEG} \cite{ni_2024_survey}. In contrast, research dedicated to \ac{EMG} \acp{FM} remains limited.

MOMENT \cite{goswami_2024_moment} is a 350-M parameter FM pre-trained on a wide range of domains and temporal resolutions.
While the original work did not consider EMG, recent works \cite{chen_2025_waveformer, chen_2025_physiowave} evaluated it for hand-gesture classification on the EPN612\cite{benalcazar_emg-epn-612_2020}, NinaPro DB5\cite{ninapro_db5}, NinaPro DB6\cite{ninapro_db6}, and UCI EMG\cite{emg_data_for_gestures_481} datasets.

PhysioWave \cite{chen_2025_physiowave} introduced a modality-specific FM for EMG, 
primarily evaluated on hand-gesture classification using the NinaPro DB5, EPN612, and UCI EMG. PhysioWave achieved a performance comparable to MOMENT with a much smaller parameter count (5M for the smaller model).

WaveFormer \cite{chen_2025_waveformer} further refined this approach with a lightweight Transformer model (3.1 M parameters) tailored for sEMG-based gesture recognition.
WaveFormer improved performance over PhysioWave on the EPN612, NinaPro DB5, NinaPro DB6, and UCI EMG datasets.

Jiang et al. \cite{jiang_2025_towards} introduced PhysioOmni, a multi-modal FM covering \ac{EEG}, \ac{ECG}, \ac{EOG}, and \ac{EMG}, based on the same encoder as in LaBraM \cite{jiang_2024_labram} ($5.8$M parameters per modality, considering its smallest-Base version). They adopt a decoupled multimodal tokenizer and masked signal modelling to ensure robustness under arbitrary missing modalities during inference. During pre-training, PhysioOmni relies on a fixed set of modalities (EEG, EOG, and EMG), which makes the architecture unsuitable for pre-training when one modality is missing. Downstream evaluations include sleep stage classification  \cite{alvarez_2021_hmc} and gait kinematic regression\cite{brantley_2018_fbm}. 

In this context, existing EMG FMs are limited by the small number of explored downstream tasks, as well as by the lack of deployability on embedded platforms. TinyMyo is the first EMG FM to cover a broader range of downstream tasks with \ac{SoA} performance, as will be extensively described in Section \ref{sec:Results}, while also being deployed on an ultra-low-power \ac{MCU}.

\section{Methods}\label{sec:Methods}
\begin{figure*}[t]
    \centering
    \includegraphics[width=1.35\textwidth]{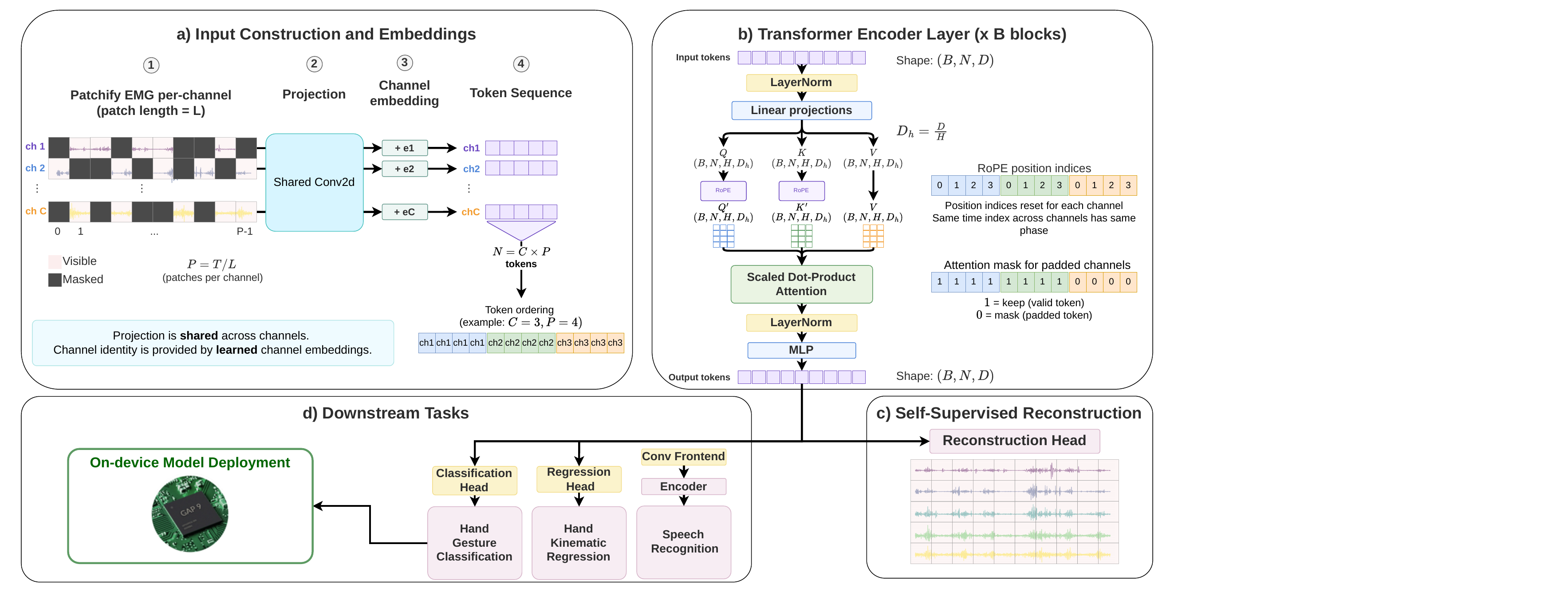}%
\caption{Overview of the TinyMyo architecture. 
\textbf{(a) Input construction and embeddings:} the multi-channel EMG signal is patchified independently for each channel, projected using a shared convolutional patch embedder, combined with learned channel embeddings, and flattened into a channel-wise token sequence. During pre-training, a subset of patches is randomly masked. 
\textbf{(b) Transformer encoder:} the token sequence is processed by $B$ bidirectional Transformer encoder blocks ($B=4$ for TinyissimoMyo and $B=8$ for TinyMyo). RoPE positional encoding is applied with position indices reset independently for each channel, so that tokens corresponding to the same temporal index across channels share the same positional phase. Optional attention masking is used for padded channels. 
\textbf{(c) Self-supervised reconstruction:} during pre-training, the encoder output is passed to a lightweight reconstruction head to reconstruct the masked EMG patches. 
\textbf{(d) Downstream tasks and deployment:} after pre-training, the reconstruction head is removed and replaced by task-specific heads. For hand gesture classification and hand kinematic regression, linear classification or regression heads are used. For speech recognition, the EMG input is first processed by a convolutional frontend and then passed through the pre-trained Transformer encoder followed by a task-specific head. The hand gesture classification model is used as the representative downstream task for on-device deployment on the GAP9 microcontroller.}
    \label{fig:architecture}
\end{figure*}

Figure \ref{fig:architecture} shows the TinyMyo framework. 
In the following sections, we present the strategy adopted for model pre-training (Sect.~\ref{sec:pretrain}) and downstream tasks (Sect.~\ref{sec:downstream}).  

\subsection{Foundation Model Pre-Training}\label{sec:pretrain}

\subsubsection{Tokenization}
We adopt a channel-independent patching strategy, preserving per-channel granularity at the embedding stage.
Unlike \ac{EEG} models that mix spatial information during tokenization via 2D convolutions spanning multiple channels \cite{tegon_2025_femba} or project them into a unified latent space \cite{doner2025luna}, we embed channels independently using a $1\times L$ kernel, and the subsequent transformer layers learn inter-channel dependencies via attention~\cite{dimofte_2025_cerebro}.
This design choice is grounded in the physiological nature of EMG recordings, where each channel corresponds to a distinct anatomical location.

Formally, given an input EMG signal $\mathbf{X} \in \mathbb{R}^{T \times C}$, with $T$ timesteps and $C$ channels, we apply a sliding window with patch length $L$ and stride $S$ independently to each channel. This yields a grid of patches $\mathbf{P} \in \mathbb{R}^{C \times N_p \times L}$, where $N_p = \lfloor (T - L) / S \rfloor + 1$ is the number of temporal patches per channel.

We employ a window of $T = 1000$ timesteps, $L = S = 20$, and $C = 16$, yielding to $50$ tokens per channel ($N_p$) and a total sequence length of $800$ tokens ($N$). 

As input sequence for the Transformer, we flatten the channel and temporal dimensions, resulting in a total sequence length of $N = C \times N_p$. Each patch $\mathbf{P}_{c,i}$ ($c$ is the channel index and $i$ is the patch index) is mapped to a latent embedding of dimension $d_e$ via a \textit{shared} learnable linear projection $\mathbf{W}_{\mathrm{proj}} \in \mathbb{R}^{d_e \times L}$. The resulting token sequence $\mathbf{E} \in \mathbb{R}^{N \times d_e}$ consists of embeddings
 $\mathbf{E}_{k} = \mathbf{W}_{\mathrm{proj}} \, \mathbf{P}_{c,i}^{\top}$
where the token index $k$ corresponds to the flattened arrangement of channel-time pairs $(c,i)$. We do not add learnable positional embeddings; instead, we inject positional information via Rotary Position Embeddings (RoPE)~\cite{su_2023_roformerenhancedtransformerrotary} within the attention blocks. To account for the multi-channel structure of EMG signals, we adopt a channel-wise temporal indexing scheme for positional encoding. When tokens are flattened across channels, assigning a single continuous sequence of positional indices would incorrectly impose a temporal ordering between channels. Instead, we reset the positional indices at the boundary of each channel (e.g., $[0 \cdots P-1; 0 \cdots P-1; \dots]$). As a result, tokens corresponding to the same time step across different channels share identical rotary positional encodings. This preserves temporal alignment while avoiding spurious inter-channel ordering, enabling the model to capture synchronous cross-channel activation patterns.

\subsubsection{Architecture}\label{subsec:pretre_archi}
We adopt an Encoder-Decoder architecture. 

The Encoder design (see Fig.~\ref{fig:architecture}) takes inspiration from DeiT-Ti~\cite{pmlr-v139-touvron21a}, a lightweight vision transformer introduced in the image domain. The original DeiT-T configuration consists of 12 transformer layers. In our work, we train two architectures, with 8 (TinyMyo) and 4 (TinyissimoMyo) layers, respectively; such a reduction aims to prioritize efficiency for edge deployment. We adopt an embedding dimension of 192 and 3 attention heads, as in previous work. We utilize an MLP ratio of 4.0 (size 768), QKV bias (without QK norm), and a dropout rate of 0.1 for attention, projection, and drop-path. The lightweight Decoder shares the embedding dimension of 192.

We adopt a RoPE approach,  which introduces position information by rotating query and key vectors in 2D planes instead of adding a position vector; such a method has already been adopted for biosignals \cite{wang_2024_eegpt, chen_2025_physiowave, chen_2025_waveformer}. 

During pre-training, the Encoder's output is passed to a lightweight Decoder (3.9k parameters), consisting of a linear layer that outputs a sequence $\mathbf{\hat{P}}$, which is a reconstruction of the original patch sequence $\mathbf{P}$. Our approach draws inspiration from SimMIM~\cite{xie2022simmimsimpleframeworkmasked}, which demonstrates that a \textit{minimal} reconstruction head (e.g., a single linear layer) is sufficient for effective self-supervised pre-training.

The use of a lightweight decoder is intentional: placing the full reconstruction burden on the encoder forces it to learn semantically rich and task-agnostic representations, rather than relying on a high-capacity decoder to compensate for missing information. This design strategy is supported by recent survey analyses in the masked image modeling literature: shallow decoders limit shortcut learning and encourage the formation of stronger latent representations within the encoder~\cite{zhang2022surveymaskedautoencoderselfsupervised}. As a result, the learned features exhibit improved generalization across downstream tasks and recording conditions.

\subsubsection{Pre-training objectives}
We adopt a random-masking approach \cite{xie2022simmimsimpleframeworkmasked}, where a random subset $\mathcal{M}$ of tokens is replaced by a learnable $\mathcal{[MASK]}$ token.
Applying the masking over each patch enforces contextual inference over tens of milliseconds (physiologically meaningful burst segments), rather than trivial gap filling. 
We adopt a masking ratio of $50\%$, as in previous works~\cite{dimofte_2025_cerebro, tegon_2025_femba} which have already shown promising performances on biosignals. 

We employ the Smooth L1 Loss function as reconstruction
objective between masked and original patches, as in \cite{chen_2025_physiowave}. We define the following loss components:
\begin{equation}
    \mathcal{L}_{\mathrm{masked}} = \frac{1}{|\mathcal{MASK}|} \sum_{(c,i) \in \mathcal{MASK}} \mathrm{SmoothL1}(\mathbf{P}_{c,i}, \mathbf{\hat{P}}_{c,i}),
\end{equation}
\begin{equation}
    \mathcal{L}_{\mathrm{visible}} = \frac{1}{|\mathcal{M}|} \sum_{(c,i) \in \mathcal{M}} \mathrm{SmoothL1}(\mathbf{P}_{c,i}, \mathbf{\hat{P}}_{c,i}),
\end{equation}
where $\mathcal{MASK}$ and $\mathcal{M}$ are the sets of masked and visible patches, respectively. To avoid overfitting on unmasked input, the loss on visible patches is reduced by a factor of $0.1$. The final reconstruction objective is given by:
\begin{equation}
    \mathcal{L}_{\mathrm{total}} = \mathcal{L}_{\mathrm{masked}} + 0.1 \cdot \mathcal{L}_{\mathrm{visible}}
\end{equation}

Reconstruction fidelity in the time domain is quantified on the validation set using the \ac{RMSE} and the \ac{R} between the reconstructed and reference EMG signals. In the frequency domain, reconstruction fidelity is assessed using \ac{PSD} analysis; the \ac{RMSE} and the \ac{R} of the \acp{PSD} between original and reconstructed signals are computed.
All experiments are implemented in Python 3.10 using PyTorch Lightning and Hydra for modularity, configurability, and reproducibility. The proposed foundation model is
trained on the CSCS Alps and Cineca Leonardo HPCs using 4x NVIDIA GH200 and 4x NVIDIA A100 GPUs respectively in Distributed Data Parallel (DDP) mode. We adopt the AdamW optimizer ($\beta_1=0.9, \beta_2=0.98$, weight decay 0.01) in \textit{bfloat16-mixed precision} with a global batch size of 2048 (512 per GPU in DDP mode), utilizing gradient clipping threshold of 1. The learning rate follows a cosine schedule for 30 epochs (3 warm-up epochs), peaking at $5\times 10^{-4}$ (minimum $1\times 10^{-6}$). The model processes sequences of 1000 timesteps with a masking ratio of 0.50 .

\subsubsection{Pre-training Datasets}
\textit{Ninapro DB6} \cite{ninapro_db6} contains sEMG from 10 subjects performing 7 hand grasp types, each repeated 12 times over 5 separate recording days. Signals were acquired with Trigno sensors (14 channels, $2~\text{KHz}$ sampling rate), placed on the forearm. The total dataset size is $20.3~\text{GB}$. 

\textit{Ninapro DB7}\cite{ninapro_db7} contains data from 20 intact subjects and 2 transradial amputees, performing 40 movements spanning isolated finger/wrist actions and grasp patterns. Data were collected using Trigno sensors (12 channels, $2~\text{kHz}$ sampling rate), placed on the forearm. The total dataset size is $30.9~\text{GB}$.

\textit{EMG2Pose}\cite{salter_emg2pose_2024} contains data from 193 participants, performing 29 different hand motions. Data were recorded with a custom wristband (16 channels, $2~\text{kHz}$ sampling rate). The total dataset size is $431~\text{GB}$.

\subsubsection{Pre-training Data Processing}\label{subsect:pretrain_preproc}
EMG data are band-pass filtered ($4^{th}$ order Butterworth, $20-450~\text{Hz}$), notch filtered ($50~\text{Hz}$), normalized to $[-1, 1]$, and segmented into 1000-sample windows ($500~\text{ms}$ at $2~\text{kHz}$) with 50\% overlap, providing sufficient temporal context \cite{smith_emgwindow_2011}
while controlling computational load. 

Datasets with fewer than 16 channels are zero-padded and masked out when performing attention. Given that each training batch might contain data from each dataset, this approach ensures consistency in the number of channels per batch.

The datasets are concatenated during pre-training, without applying any explicit balancing strategy, in order to maximize the amount of data available for representation learning. While larger datasets may therefore have a stronger influence on the learned representations, the inclusion of \textit{Ninapro DB6} and \textit{DB7} increases platform variability and highlights the flexibility of the proposed architecture in handling different channel counts and acquisition settings.

We did not apply any data augmentation due to the limited effect in scaling-laws regimes~\cite{kaplan2020scalinglawsneurallanguage} and in line with previous works \cite{chen_2025_physiowave, chen_2025_waveformer}.

\subsection{Downstream Tasks}\label{sec:downstream}
\subsubsection{Pipeline Overview}
For all downstream tasks, the pre-trained encoder is retained and the decoder removed.

The Encoder output is processed through a channel-fusion and temporal pooling module (detailed in the following subsection), after which the resulting representation is fed into task-specific prediction heads, whose specifics are provided in the corresponding downstream task sections (see Sec.\ref{subsec:downstream}).

For each temporal patch index $p$, the Encoder outputs a set of channel-wise embeddings $\{\mathbf{z}_{c,p}\in\mathbb{R}^{d_e}\}_{c=1}^{C}$. We fuse these representations by concatenation, producing a per-patch vector $[\mathbf{z}_{1,p}\Vert\cdots\Vert\mathbf{z}_{C,p}]\in\mathbb{R}^{C \times d_e}$. This choice preserves electrode-specific information by retaining the per-channel structure, allowing the downstream linear head to assign independent weights to each channel. Ablation studies confirmed that concatenation consistently outperforms mean-based fusion. 

Following channel fusion, we apply temporal average pooling:$\
    \mathbf{h_{c}}=\frac{1}{N_c,p}\sum_{p=1}^{N_p}\mathbf{z}_{p},$
    where $\mathbf{z}_{p}\in\mathbb{R}^{C \times d_e}$. 
This aggregation step yields a compact representation while preserving discriminative spatiotemporal structure, and it serves as the input to the final linear task-specific head.

\subsubsection{Downstream Tasks}\label{subsec:downstream}
\begin{table}[t]
\centering
\caption{Summary of Downstream Datasets.}
\label{tab:downstream_dataset}
\scriptsize
\setlength{\tabcolsep}{3.5pt} 

\begin{tabular}{l c c c c l}
\toprule
\textbf{Dataset} & \textbf{\#S} & \textbf{Hz} & \textbf{Ch} & \textbf{Loc.} & \textbf{Metric} \\
\midrule
\multicolumn{6}{l}{\textit{Hand Gesture Classification}} \\
Ninapro DB5 [14] & 10 & 200 & 16 & FA & Acc, F1 \\
EPN-612 [15]     & 612 & 200 & 8 & FA & Acc, F1 \\
UCI-EMG [16]     & 36 & 200 & 8 & FA & Acc, F1 \\
GNI [5]          & 100 & 2k & 16 & H/A & CLER \\
\midrule
\multicolumn{6}{l}{\textit{Hand Kinematic Regression}} \\
Ninapro DB8 [17] & 12 & 2k & 16 & FA & MAE \\
\midrule
\multicolumn{6}{l}{\textit{Speech Production \& Recognition}} \\
Gaddy [18]       & 1 & 1k & 8 & F/N & WER \\
\bottomrule
\end{tabular}

\vspace{2mm}
\begin{minipage}{\columnwidth} 
\scriptsize
\emph{Loc: Location (FA: Forearm, H/A: Hand/Arm, F/N: Face/Neck). \\
Pre-processing and windowing details are provided in Section III-B.}
\end{minipage}

\end{table}
The pretrained encoder is evaluated on the datasets listed in Table \ref{tab:downstream_dataset}. We classify the downstream tasks into three categories: Hand Gesture Classification, Hand Kinematic Regression, and Speech Synthesis and Recognition.

\textbf{Downstream 1: Hand Gesture Classification}

This task consists of classifying different types of hand-wrist movements and is evaluated on the following datasets:

\begin{itemize}
\item \textit{Ninapro DB5 \cite{ninapro_db5}}: data from 10 intact subjects, using two Thalmic Myo Armbands (16 channels, $200~\text{Hz}$). 
The task consists of classifying 52 hand-wrist movements.

\item \textit{EPN-612 \cite{benalcazar_emg-epn-612_2020}} data from 612 subjects, using the Myo Armband (8 channels, $200~\text{Hz}$). 
The task consists of classifying 5 hand movements. 

\item \textit{UC Irvine \cite{emg_data_for_gestures_481}} data from 36 subjects (8 channels, $200~\text{Hz}$). The task consists of classifying 7 movements.
\end{itemize}

Datasets are pre-processed using a band-pass filter (Butterworth, $4^{th}$ order, $[20-90]~\text{Hz}$), followed by a notch filter ($50~\text{Hz}$). 
Then, windowing is applied. We perform an ablation over two different window sizes: $200$ samples with 25\% overlap ($1$~s at $200$~Hz), and $1000$ samples with 25\% overlap ($5$~s at $200$~Hz, in line with \cite{chen_2025_physiowave, chen_2025_waveformer}). Finally, each dataset is normalized using per-channel \textit{z}-score normalization.

Compared to the pre-training, datasets with fewer than 16 channels are not zero-padded because tokenization is performed on a per-channel basis: having fewer channels simply results in fewer tokens, eliminating the need for padding.

For Ninapro DB5, we perform a per-repetition split as in \cite{chen_2025_physiowave}, each recording contains six repetitions, and for each subject, we assign whole repetitions to a single split so that no subject data overlaps across splits (repetitions 1-3-4-6 are used for training, rep. 2 for validation, and rep. 5 for testing). We apply several data augmentation techniques to the training set, including amplitude scaling (with a scaling factor between 0.9 and 1.1), time jitter (0.01 jitter ratio, indicating how the noise is scaled relative to each channel’s standard deviation), and channel dropout (with a dropout probability of 0.05). For every training example, three distinct augmented variants are generated and combined with the original sample.

The UCI EMG dataset is split by subject index into three disjoint groups: subjects 1--24 for training, subjects 25--30 for validation, and subjects 31--36 for testing. Only subjects 11 and 30 performed the ``extended palm" gesture, so they were excluded from all data splits,  hence reducing the total number of gestures to 6.

For EPN-612, we preserve the original per-user organization \cite{benalcazar_emg-epn-612_2020} and derive splits per user. When constructing the testing split, held-out examples per user are used, ensuring that samples from the same user do not appear across train, validation, or test splits.

The model head architecture is lightweight and consists of a single fully connected linear layer that maps the pooled feature representation (max $192\times N_c$, $N_{c}=$ number of channels ) to the final class logits ($N_{o}$, $N_{o}=$number of outputs). No additional hidden layers, normalization modules, or non-linearities are used in the head. Hence, the model head has a maximum of $192 \times N_{c} \times N_{o}$ parameters. 
For training, we employ the cross-entropy loss computed over the predicted logits and the corresponding ground-truth labels.
We evaluate performance using Accuracy, F1-score, and AUROC.

We repeat each experiment five times using five distinct random seeds and report the mean performance across runs.

An additional dataset is also considered as part of this downstream task:
\begin{itemize}
\item \textit{Generic Neuromotor Interface} \cite{generic_neuromotor_interface_2025} contains data collected from 4900 participants, recorded with a custom wrist-band developed by Meta Reality Labs (16 channels, $2\text{kHz}$). Data from only 100 subjects are publicly available \cite{metagit}, randomly selected from the entire cohort. 
\end{itemize}

Specifically, we tackle the \textit{discrete-gesture} task presented in the original work, i.e., classifying  9 fine finger movements. We treat this task separately, compared to the three other datasets presented above, owing to substantial methodological differences in data processing and evaluation protocols. 

No pre-processing is applied to the data. Signals are windowed using $8000~\text{ms}$ windows, no overlap. 
The train dataset contains 80\% of data from each subject for train, 10\% for validation, 10\% for test.
The model head consists of a single linear layer, with input size of $N_{c} \times 192$, output size $9$.  We employ the Cross-Entropy Loss function, and we conduct 5 runs with 5 seeds values. 

We evaluate the performance with the Classification Error Rate (CLER), defined as \textit{the proportion of ground-truth labels for which the matching prediction is incorrect} \cite{generic_neuromotor_interface_2025}. The CLER is computed per-gesture and reported as the average across gestures. The CLER formula is given by: 
\[
\mathrm{CLER} =
\frac{1}{|N_{G}|}
\sum_{g \in G}
\left(
1 - \frac{C_{g,g}}{\sum_{p \in G} C_{g,p}}
\right).
\]
where $N_{G}$ is the total number of gestures, $g \in N_{G}$, $C_{g,g}$ is the number of correct predictions for gesture $g$ and
$\sum_{p \in G} C_{g,p}$ is the total number of ground-truth occurrences of $g$.
In the original work, the architecture is tested on the full sequence of EMG data.  The architecture proposed in this work cannot process the entire sequence due to quadratic complexity. We adopt a windowed inference approach \label{windowed_inf}, which involves sliding a window of $8\text{s}$, with $2\text{s}$ stride. In order to compare the original LSTM architecture with the proposed Transformer model, the same windowed inference strategy is applied to the LSTM during evaluation, and both results are reported.

\textbf{Downstream 2: Hand Kinematic Regression} 

This task consists of predicting continuous hand and wrist movement (regression). We employ the \textit{Ninapro DB8}~\cite{ninapro_db8} dataset, containing EMG data from 12 subjects (16 channels, $2~\text{kHz}$), with 3 sessions per subject. Data were collected with an instrumented glove from ground-truth.
The task consists of regressing 5 Degrees of Freedom (DoFs), defined as linear combinations of the 18 DoFs of the instrumented glove. 

Each EMG channel is independently standardized using \textit{z}-score normalization. No filtering is applied since the original data is already released filtered ($4\textsuperscript{th}$ order bandpass Butterworth filter, $[10-500 ~\text{Hz}]$. The glove data undergo a mapping from 18 DoFs to 5 Degrees of Actuation. Both EMG and glove data are then segmented into fixed-size windows with no overlap in two different settings: $200$ samples and $1000$ samples ($100$ and $500$~\text{ms} respectively at $2$~kHz).

Following the author's recommendations, we consider the first two sessions for training, the third as validation.
The model head architecture (788 K parameters) consists of a compact series of pointwise and depthwise convolutions, followed by upsampling to a fixed length. We employ the L1 loss function. 
We report the mean absolute error (MAE), root mean squared error (RMSE),
and the coefficient of determination $R^2$, averaged over 5 runs with 5 random seed values.

\textbf{Downstream 3: Speech Synthesis and Recognition}

We employ the open-vocabulary dataset presented by Gaddy et al. \cite{gaddy-klein-2020-digital}. 
It comprises 8 channels, sampled at $1000~\text{Hz}$, recorded using wet electrodes, placed on the face.
Data were collected from one subject in multiple sessions while reading pieces of public domain books. The dataset contains both \textit{vocalized} (i.e., normal articulation with sound production) and \textit{silent} (i.e., articulation without sound production) EMG data. In the vocalized setting, EMG data were recoded together with a microphone, serving as ground truth. For the silent setting, audio recorded during vocalized experiments is aligned to the corresponding silent EMG data using the dynamic time warping algorithm \cite{gaddy-klein-2020-digital}.

For both tasks, we adopt the data processing pipeline of \cite{gaddy-klein-2021-improved}. A cascade of notch filters is applied to suppress the first six harmonics of power line frequency ([$60, 240, ..., 420]~\text{Hz}$), followed by a high-pass filter (Butterworth, $2~\text{Hz}$, 3rd order) and downsampling from to $516.79~\text{Hz}$ \cite{gaddygit}.

For this dataset, we tackle two sub-tasks: \textit{Speech Synthesis} and \textit{Speech Recognition}. \textit{Speech Synthesis} aims at producing audio from EMG data. 
The full speech production pipeline consists of three steps: a Transduction Model, a Vocoder, and an Automatic Speech Recognition (ASR) model. First, a Transduction model is trained to reconstruct audio target features (Mel-Frequency Cepstral Coefficients, MFCC). It consists of 3 residual blocks (as in \cite{gaddy-klein-2021-improved}) that downsample the input (1600 samples to 200 samples). The 200 samples are the input to the pre-trained Transformer Encoder; we add relative positional embeddings, following \cite{gaddy-klein-2021-improved}. A lightweight linear head follows, with an input dimension of $192 \times N_{c}$, output dimension of 26 MFCC.  

The model is trained using as loss function the mean of pair-wise MFCC distances ($L$) plus an auxiliary phoneme loss ($L_{phon}$) ~\cite{gaddy-klein-2021-improved}. For silent EMG data, DTW is used to align the MFCC features predicted from silent EMG to the sequence of target MFCC features obtained from a parallel vocalized recording. 
Predicted MFCC audio features are fed to a HiFi-GAN vocoder, which converts the audio features into raw audio waveforms. We adopt the publicly available pre-trained HiFi-GAN vocoder \cite{gaddygit} without any fine-tuning.

For model evaluation, we employ the Wav2Vec2 \cite{baevski_2020_wav2vec} ASR model, pretrained on LibriSpeech (English Language) within SpeechBrain \cite{speechbrain}. 

\textit{Speech Recognition}
consists of recognizing speech (words) without any acoustic output, yielding text as output.
Direct EMG-to-text prediction avoids intermediate audio synthesis and reduces model complexity, yielding improved recognition accuracy \cite{gaddy_2022_phdthesis}.
This makes the approach more suitable for deployment on resource-constrained embedded platforms.

The model architecture is the same as before.
The input to the final linear layers has a dimension of $N_{c} \times 192$, the output dimension is 37 (26 lowercase English letters, 10 digits, and space character). We adopt the Connectionist Temporal Classification (CTC) \cite{graves_2006_connectionist} loss function with a 4-gram language model trained on the LibriSpeech dataset \cite{LibriSpeech}. 
During inference, beam search is applied to explore the search space and select the paths with the highest likelihood. 

The dataset is split as in \cite{gaddy-klein-2021-improved}. Training is performed using both silent and vocalized EMG recordings; testing is performed on silent data.
Data are split into training, validation, and test sets using the original index file containing pairs of book and sentence indexes. Each utterance is assigned to one of the three splits. During training, a size-aware sampler is used to group examples, together producing compact batches and a more efficient training stage.

For both tasks, we use the Word Error Rate (WER) as the evaluation metric, defined as
$\text{WER} = \frac{S + D + I}{N}$,
where $S$, $D$, and $I$ denote the number of substitution, deletion, and insertion errors, respectively, and $N$ is the total number of words in the reference transcription.

\subsubsection{Experimental Settings}

All downstream experiments use the AdamW optimizer ($\beta_1=0.9, \beta_2=0.98$) with a batch size of 32, weight decay 0.01, and label smoothing 0.1. We employ a cosine learning rate schedule (50 epochs, 5 warm-up) peaking at $5\times 10^{-4}$ (min $1\times 10^{-5}$) with 0.90 layer-wise decay and 0.1 drop-path. Early stopping is adopted with a patience of 7 epochs.

We evaluate the downstream tasks under two different settings: \textit{(i)} \textit {Full-supervised (FS)}, where the model is trained from scratch on the task-specific dataset without relying on pre-trained weights; (ii) \textit{Full Finetuning (FT), where the entire model is fine-tuned using layer-wise learning rate decay to avoid catastrophic forgetting} \cite{kenneweg_forgetting_22}.
\subsection{Model Deployment}\label{sec:methods_deply}
We deploy the TinyMyo on the GreenWaves Technologies GAP9~\cite{greenwaves_gap_sdk} processor, a multi-core RISC-V MCU designed for ultra-low-power edge AI. The architecture features a hierarchical memory organization consisting of a large off-chip L3 (HyperRAM), a 1.5 MB on-chip L2 (shared memory), and a 128 kB L1 (Tightly Coupled Data Memory) shared among a cluster of 8 compute cores and a dedicated cluster controller.

To enable the execution of the Transformer-based architecture, which exceeds the on-chip L2 capacity in terms of parameters and intermediate activation maps, we developed a custom deployment toolchain that implements a hierarchical streaming architecture.

Unlike traditional MCU inference engines that require the full model to reside in on-chip memory, we treat L2 as a streaming buffer. We implement a multi-level tiling strategy:
\begin{itemize}
    \item \textbf{L3} $\rightarrow$ \textbf{L2}: Large tensors (weights and activations) are stored in external L3 memory. They are retrieved into L2 slabs on demand. For example, in linear and multi-head self-attention layers, weights are streamed slab by slab to prevent memory overflows.
    \item \textbf{L2} $\rightarrow$ \textbf{L1}: Computation is performed by moving data from the L2 slabs into L1 tiles. The cluster controller orchestrates high-bandwidth DMA transfers, while the 8 worker cores execute compute kernels in parallel.
\end{itemize}
To minimize the overhead of external memory access, we implement a double-buffered pipeline. While the worker cores compute the current slab (N), the cluster controller asynchronously pre-fetches the next slab (N+1) from L3 to L2 or from L2 to L1. This effectively hides the latency of data movement behind the compute-bound operations of the transformer blocks.

We apply \texttt{INT8} quantization to convert all weights and activations from \texttt{FP32} to 8-bit fixed-point representations. 
All intermediate tensors, including those in MHSA (with integer softmax) and GELU/LayerNorm, are stored and processed in \texttt{INT8} format, with only the LayerNorm scale and shift parameters ($\gamma$ and $\beta$) retained in \texttt{FP32}. The only remaining \texttt{FP32} operation is the final classification layer, which consumes \texttt{INT8} inputs and weights but accumulates to \texttt{FP32} logits to enable a direct comparison with the original \texttt{FP32} baseline; in principle, this layer could also be executed in \texttt{INT8}.

To maximize the utilization of the limited 1.5 MB L2 memory, we employ an offline liveness analysis pass. Instead of dynamic allocation, the deployment flow calculates a static memory arena. Tensors are assigned offsets within this arena based on their execution lifetime, allowing mutually exclusive buffers (e.g., non-overlapping intermediate feature maps) to share the same physical memory addresses.

To maintain numerical fidelity while avoiding costly floating-point operations, all transformer primitives are implemented in fixed-point arithmetic\cite{kim_2021_bert}. We use integer-only variants of softmax (i-Softmax), layer normalization (i-LayerNorm), and GELU activation (i-GELU), which leverage lookup tables and multiplicative inverses to approximate their floating-point counterparts with minimal error.

As proof of concept, we deploy the \textit{Hand Gesture Classification} downstream task on the \textit{EPN-612 dataset}, for both the TinyMyo and TinyissimoMyo model sizes.
\section{Results}
\label{sec:Results}
\subsection{Pre-training}
\begin{figure}
    \centering
    \includegraphics[width=1\linewidth]{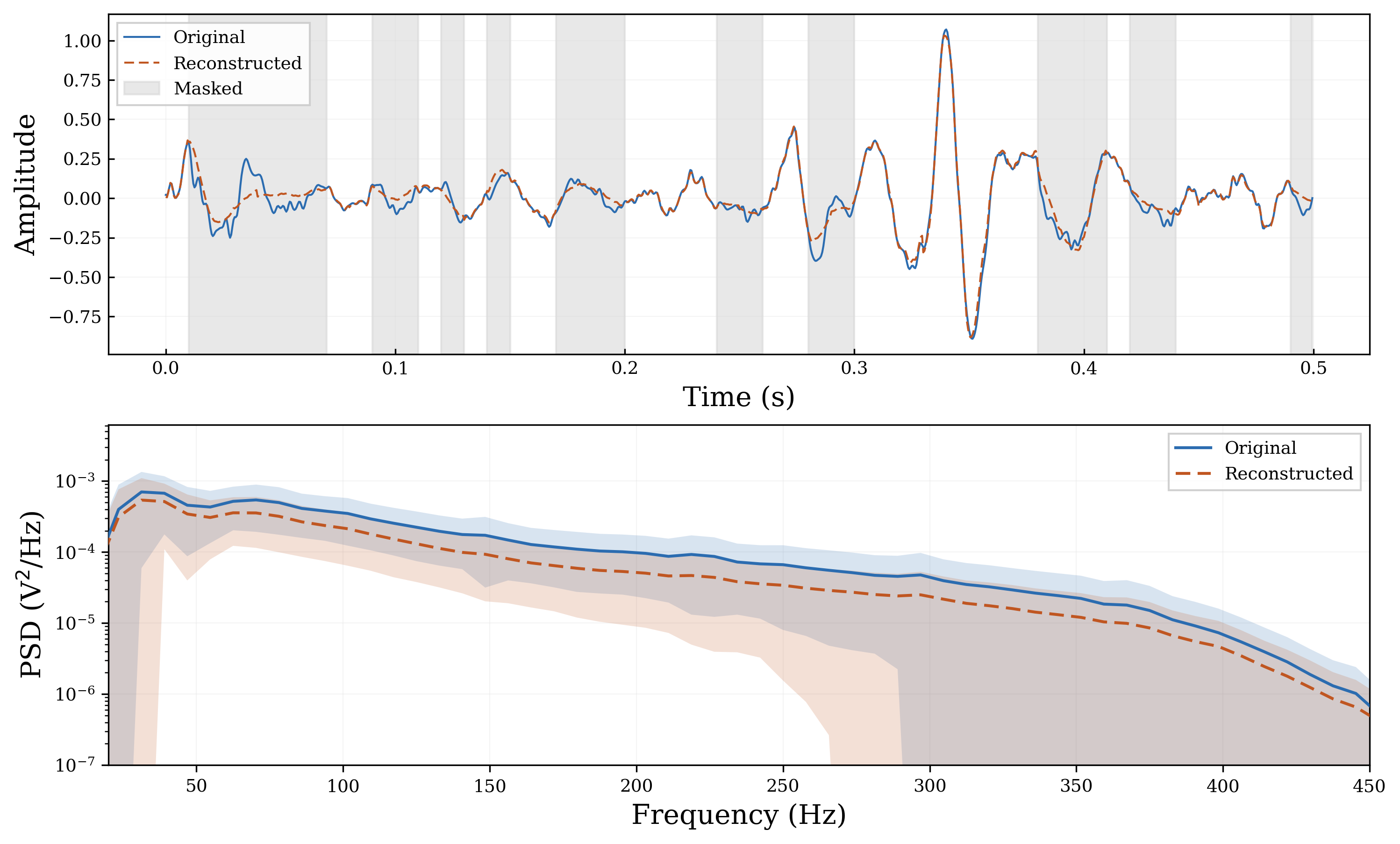}
    \caption{Top: Example of reconstruction for an EMG window by TinyMyo. Bottom: Example of PSD computed on the original and reconstructed frames across the entire dataset (validation set), within the 20-450 Hz range, 50\% masking ratio. Data correspond to the Ninapro DB6 dataset.}
    \label{fig:psd_datasets}
\end{figure}

\begin{table}[h]
\vspace{-0.4cm}
\centering
\caption{Reconstruction performance (RMSE, $R$) in time and frequency domains.}
\label{tab:time_freq_recons}
\setlength{\tabcolsep}{3pt}
\renewcommand{\arraystretch}{1.1}
\begin{tabular}{l ccc ccc}
  \toprule
  \textbf{Dataset} & \multicolumn{2}{c}{\textbf{Time}} & \multicolumn{2}{c}{\textbf{Freq.}} \\
  \cmidrule(lr){2-3} \cmidrule(lr){4-5}
  & \textbf{RMSE} & \textbf{$R$} & \textbf{RMSE} & \textbf{$R$}\\
  \midrule
  Ninapro DB6 ~\cite{ninapro_db6} 
    & $1.5 \times 10^{-1}$ & 0.804 & $7.8\!\times\!10^{-5}$ & 0.992\\
  Ninapro DB7 ~\cite{ninapro_db7} 
    & $1.5 \times 10^{-1}$ & 0.813 & $7.8\!\times\!10^{-5}$ & 0.993\\
  EMG2Pose~\cite{salter_emg2pose_2024} 
    & $1.2 \times 10^{-1}$ & 0.809 & $5.4\!\times\!10^{-5}$ & 0.990\\
  \bottomrule
\end{tabular}
\vspace{-0.3cm}
\end{table}
Table \ref{tab:time_freq_recons} reports the RMSE and $R$ achieved on the validation set for the three datasets used for pre-training. 
Figure \ref{fig:psd_datasets} shows a representative example of reconstruction in both time and frequency domains for one of the datasets used for pre-training.

With a parameter count of only 3.6 million (with only 0.1\% of parameters for the decoder), in line with previous works \cite{chen_2025_waveformer}, our model demonstrates that high reconstruction fidelity can be maintained even under a strict parameter budget and low computational requirements 
In terms of computational complexity, the model requires 4.0 GFLOPs for pre-training. Inference costs vary by downstream task complexity: approximately 1.1 GFLOPs for speech tasks, 1.9 GFLOPs for EPN-612/UCI-EMG, 4.8 GFLOPs for Ninapro datasets, and 13.6 GFLOPs for the Generic Neuromotor.

\begin{table*}[t]
\centering
\scriptsize
\caption{Hand Gesture Classification Task — Comparison with State of the Art.}
\label{class_res}

\renewcommand{\arraystretch}{1.15}
\setlength{\tabcolsep}{1.5pt}
\begin{tabular}{l | c | ccc | ccc | ccc }
\toprule
\textbf{Model} & \textbf{Params} &
\multicolumn{3}{c|}{\textbf{NinaPro DB5}} &
\multicolumn{3}{c|}{\textbf{EPN-612}} &
\multicolumn{3}{c}{\textbf{UCI-EMG}} \\
\midrule
& & Acc & F1 & AUROC  & Acc & F1 & AUROC & Acc & F1 & AUROC \\
\midrule

Moment (as in \cite{chen_2025_physiowave}) 
& 385M 
& 86.41 & 74.42 & N.A. 

& 90.87 & 90.16 & N.A. 
& 90.45 & 91.75 & N.A. \\

Moment (as in \cite{chen_2025_waveformer}) & 
& 86.41$\pm$1.13 & 76.84$\pm$1.95 & 99.34$\pm$0.12

& 93.83$\pm$0.76 & 93.75$\pm$0.82 & 99.63$\pm$0.09
& 92.79$\pm$0.64 & 91.24$\pm$0.71 & 99.47$\pm$0.11 \\
\midrule

PhysioWave-Small \textsuperscript{b} \cite{chen_2025_physiowave} & 5M
& 84.78 & 72.54 & N.A.

& 93.12 & 93.40 & N.A.
& 90.35 & 89.51 & N.A. \\
\midrule

WaveFormer\textsuperscript{a, b}\cite{chen_2025_waveformer} & 3.1M
& 87.53$\pm$0.92 & 74.66$\pm$1.78 & 99.35$\pm$0.10

& 95.21$\pm$0.68 & 95.22$\pm$0.71 & 99.70$\pm$0.07
& 93.10$\pm$0.58 & 93.20$\pm$0.62 & 99.60$\pm$0.08 \\
\midrule
TinyMyo (FS), 1 sec & 3.6M
& 83.12$\pm$0.58 & 61.72$\pm$1.63 & 97.57$\pm$0.3
& 72.97$\pm$0.1 & 72.97$\pm$0.14 & 93.38$\pm$0.11
& 93.13$\pm$0.44 & 93.13$\pm$0.46 & 99.57$\pm$0.05\\
TinyMyo (FT), 1 sec & 3.6M
& \textbf{87.98$\pm$0.96} & 74.02$\pm$2.5 & 98.06$\pm$0.18
& 75.82$\pm$0.36 & 75.9$\pm$0.31 & 94.39$\pm$0.07
& 93.76$\pm$0.59 & 93.74$\pm$0.6 & 99.57$\pm$0.09\\
TinyMyo (FS), 5 sec & 3.6M
& 83.68$\pm$0.43 & 62.99$\pm$1.25 & 97.34$\pm$0.3
& 95.1$\pm$0.07 & 95.09$\pm$0.07 & 99.54$\pm$0.01
& 96.93$\pm$1.08 & 96.91$\pm$1.08 & 99.92$\pm$0.02\\
TinyMyo (FT), 5 sec & 3.6M
& 87.62$\pm$0.93 & 73.1$\pm$2.46 & 98.02$\pm$0.29
& \textbf{96.57$\pm$0.15} & 96.57$\pm$0.15 & 99.61$\pm$0.06
& 96.51$\pm$0.67 & 96.51$\pm$0.67 & 99.90$\pm$0.03\\

\midrule
TinyissimoMyo (FS), 1 sec & 1.9M
& 78.02$\pm$2.4 & 47.95$\pm$6.04 & 96.86$\pm$0.86
& 72.17$\pm$0.18 & 72.18$\pm$0.21 & 93.15$\pm$0.1
& 93.7$\pm$0.03 & 93.71$\pm$0.29 & 98.6$\pm$0.04\\
TinyissimoMyo (FT), 1 sec & 1.9M
& 81.19$\pm$0.22 & 56.01$\pm$0.71 & 97.73$\pm$0.26
& 72.5$\pm$0.4 & 72.57$\pm$0.37 & 93.14$\pm$0.11
& 93.18$\pm$0.35 & 93.17$\pm$0.36 & 99.55$\pm$0.05\\

TinyissimoMyo (FS), 5 sec & 1.9M
& 76.32$\pm$0.96 & 45.53$\pm$2.27 & 96.13$\pm$0.26
& 91.99$\pm$0.12 & 91.99$\pm$0.13 & 99.14$\pm$0.03
& \textbf{97.1$\pm$0.27} & 97.08$\pm$0.28 & 99.93$\pm$0.02\\
TinyissimoMyo (FT), 5 sec & 1.9M
& 78.17$\pm$0.62 & 47.58$\pm$1.44 & 97.13$\pm$0.24
& 93.19$\pm$0.22 & 93.17$\pm$0.22 & 99.31$\pm$0.03
& 95.63$\pm$0.27 & 95.61$\pm$0.28 & 99.79$\pm$0.02\\

\bottomrule
\end{tabular}

\vspace{2mm}
\begin{minipage}{\textwidth}
\footnotesize
\justifying

\noindent{\textsuperscript{a} WaveFormer uses both \ac{EMG} and accelerometer data on the NinaPro DB5 dataset. \\}\textsuperscript{b} PhysioWave \cite{chen_2025_physiowave} and WaveFormer \cite{chen_2025_waveformer} use a window size of approx 5.12 s (1024 samples, sampling rate 200 Hz)
\end{minipage}
\end{table*}

\subsection{Down-stream Tasks Results}
\subsubsection*{1. Hand Gesture Classification}
Table \ref{class_res} reports the results obtained on the \textit{Ninapro DB5}, \textit{EPN-612}, and the \textit{UC Irvine} datasets, comparing TinyMyo to previous FM works.

On the \textit{NinaPro DB5} dataset, the fine-tuned TinyMyo model trained with a window size of $1\,\text{s}$ achieves state-of-the-art performance, with an accuracy of $87.98 \pm 0.96\,\%$ and an F1-score of $74.02 \pm 2.5\,\%$. Fine-tuning the pre-trained model yields an improvement of approximately $4.9\%$ in accuracy compared to training from scratch. The performance achieved when using a window size of $5\,\text{s}$ is comparable ($87.62\pm0.93\%$).
It can also be observed that the model with 8 Transformer blocks consistently outperforms the smaller variant.

It must be noted that this dataset exhibits a substantial imbalance toward the \textit{rest} class. To ensure a fair comparison with prior work, we did not apply any rebalancing techniques during training or evaluation.
Computing a macro-averaged accuracy score reduces the accuracy to $72.05 \pm 2.07\,\%$, as expected for an imbalanced dataset.

To further investigate the trade-off between recognition performance and responsiveness, we also evaluated TinyissimoMyo on the \textit{NinaPro DB5} dataset using a shorter input window of $200~\text{ms}$. Under this setting, TinyissimoMyo achieves an accuracy of $80.91 \pm 0.44\%$ corresponding to a reduction of $0.28\%$ compared with the $1\text{s}$ configuration. This performance decrease is consistent with the expected trade-off between shorter temporal context and classification accuracy reported in previous studies \cite{sri2021classification}.

On the \textit{EPN-612} dataset TinyMyo achieves SoA results (window size of $5\text{sec}$), with a global accuracy (FT) of $96.57\pm 0.15 \%$, F1-score of $96.57\pm 0.15\%$ and AUROC of $99.61\pm 0.06\%$. As for the previous dataset, a bigger model gives better performance.

On the \textit{UCI-EMG} dataset, the best performance is obtained with TinyissimoMyo, with an input window of \(5\,\text{sec}\), achieving an accuracy of \(97.1 \pm 0.27\) \%, an F1-score of \(97.08 \pm 0.28\) \%, and an AUROC of \(99.93 \pm 0.02\) \%. The performance of the model trained from scratch is comparable to that obtained with the pre-trained TinyMyo encoder, which we attribute to the relative simplicity of this downstream task. The performance achieved by the bigger model is comparable, taking into account the seed variability.

\begin{table}[h]
    \centering
    \caption{Discrete Gesture Classification Task- Comparison with SoA}
    \label{tab:generic_descrete}
    \begin{tabular}{lccc}
        \toprule
        \textbf{Model}      & \textbf{Params} & \textbf{CLER}  & \textbf{Inference Method} \\
        \midrule
        LSTM, Kaifosh \cite{generic_neuromotor_interface_2025} \textsuperscript{(a)}           & 6.4M            & 0.1819         & Full sequence             \\
        LSTM, Kaifosh \cite{generic_neuromotor_interface_2025} \textsuperscript{(b)}         & 6.4M            & 0.1596         & Windowed                  \\
        \midrule    
        TinyMyo - FS           & 3.6M            &   0.142$\pm$0.007  & Windowed \\
        TinyMyo - FT           & 3.6M            &   0.146$\pm$0.005   & Windowed \\
        \midrule
        TinyissimoMyo - FS & 1.8M            &   0.151$\pm$0.008    & Windowed \\
        TinyissimoMyo - FT & 1.8M           &  0.153$\pm$0.01   & Windowed \\
        \bottomrule
        
    \end{tabular}

\parbox{0.95\linewidth}{
\footnotesize{
\textsuperscript{(a)}: Full sequence results obtained with methods provided by the authors (\cite{generic_neuromotor_interface_2025}) \textsuperscript{(b)}: results obtained with the same model as in \textsuperscript{a}, with the windowed-inference method described in the corresponding Methods section. FS: model trained from scratch, FT: fine-tuning pretrained
model
}
}
\end{table}

For the \textit{Generic Neuromotor Interface - discrete gesture task} \cite{generic_neuromotor_interface_2025},  
we achieve a CLER of $0.142\pm0.007$ on the model trained from scratch, which compares to the 0.1596 (obtained with our windowed inference method, see Sec. \ref{windowed_inf}) obtained with the LSTM model proposed in the original work, however with $\approx 44\%$ reduction in model size. In this case, the results achieved with the pre-trained FM are worse ($0.146\pm0.005$), which we attribute to the intrinsic nature of the problem: to emulate the sequential processing behaviour of the original stacked LSTMs, we imposed a causal mask on the attention mechanism. This constraint restricts our architecture, initially optimized for bidirectional dependency modelling, to a strictly left-to-right context, leading to a degradation in the effectiveness of the pre-trained weights when the future information on which they were originally trained is no longer available. Table \ref{tab:generic_descrete} reports a summary of the results.

We note that our work is the first to include this dataset among the downstream tasks and to propose an alternative architecture compared to the one of the original work. 
\subsubsection*{2. Hand Kinematic Regression}

Table \ref{tab:emg_regression} reports the result for the Hand Kinematic Regression task. 
\begin{table}[h]
\centering
\scriptsize
\caption{Gesture Regression Task on Ninapro DB8 - Comparison with SoA}
\label{tab:emg_regression}

\begin{tabular}{l
                @{\hspace{2pt}}c
                @{\hspace{2pt}}c
                @{\hspace{2pt}}c}
\toprule
\textbf{Model} & \textbf{Params} & \textbf{MAE$^\circ$} & \textbf{RMSE$^\circ$} \\
\midrule

TEMPONet TCN, Zanghieri~\cite{zanghieri_2021_db8}$^{*}$  
& $<$500K & 6.89 & -- \\

Event-based Lin. Reg., Zanghieri~\cite{zanghieri_event_based_db8}$^{*}$ 
& -- & 8.8$\pm$2.3 & -- \\

DNKF, Bao~\cite{BAO202188}$^{**}$
& -- & -- & 13.5 \\
\hline

TinyMyo - FS, 100 ms ($^{***}$)
& 3.8M & 9.48$\pm$0.29 & 15.48$\pm$0.46  \\

TinyMyo - FT, 100 ms
& 3.8M & 9.05$\pm$0.11 & 14.73$\pm$0.17  \\

TinyMyo - FS, 500 ms
& 3.8M & 9.35$\pm$0.26 & 15.18$\pm$0.42 \\

TinyMyo - FT, 500 ms
& 3.8M & 8.8$\pm$0.23 & 13.15$\pm$2.44 \\\hline

TinyissimoMyo - FS, 100 ms ($^{***}$)
& 2.6M & 9.84$\pm$0.21 & 16.37$\pm$0.34 \\

TinyissimoMyo - FT, 100 ms
& 2.6M & 9.8$\pm$0.16 & 16.16$\pm$0.21 \\

TinyissimoMyo - FS, 500 ms
& 2.6M & 11.56$\pm$0.25 & 18.81$\pm$0.21 \\

TinyissimoMyo - FT, 500 ms
& 2.6M & 10.38$\pm$0.09 & 16.73$\pm$0.12 \\
\hline

\end{tabular}

\vspace{2pt}
\footnotesize
\parbox{0.95\linewidth}{
FS: trained from scratch; FT: fine-tuning\\
$^{*}$ 5 DoFs, subject-specific model; 
$^{**}$ 3 DoFs, subject-specific model; \\
$^{***}$ 5 DoFs, across-subject model; lower accuracy in the across-subject setting reflects increased task difficulty.}
\end{table}

TinyMyo (FT) achieves a MAE of $8.8^\circ\pm0.23^\circ$, with a window size of $500\text{ms}$. This value is higher than the result of~\cite{zanghieri_2021_db8} (MAE $= 6.89^\circ$), primarily due to differences in both the training and evaluation protocols. In our setting, the model is trained in an \textit{across-subject} regime using data aggregated from all participants, while prior work trained \textit{subject-specific} models, i.e., one model per individual. Learning a single model that generalizes across subjects is substantially more challenging because of inter-subject variability in muscle physiology, electrode placement, and EMG signal morphology. With a window size of $500\text{ms}$ (FT case), the bigger variant results in more than $1.5^\circ$ MAE improvement compared to the smaller one.

\subsubsection*{3. Speech Recognition}
\begin{table*}[t]
\centering
\scriptsize
\caption{Silent Speech Synthesis Task – Comparison with SoA}
\label{tab:emg-synthesis}

\begin{tabular}{l cc cc c c}
\toprule
& \multicolumn{2}{c}{\textbf{Transduction Model}} 
& \multicolumn{2}{c}{\textbf{Vocoder}} 
& \textbf{ASR} 
& \textbf{WER} \\
\cmidrule(lr){2-3} \cmidrule(lr){4-5}
\textbf{Method} 
& \textbf{Model} & \textbf{Params [M]} 
& \textbf{Model} & \textbf{Params [M]} 
&  
&  \\ 
\midrule

Transformer, Gaddy~\cite{gaddy-klein-2021-improved}
& CNN + Transformer & 54
& HiFi-GAN & 13.92  
& DeepSpeech 
& 36\% \\

Diff-ETS, Ren~\cite{ren_speech_24}
& as \cite{gaddy-klein-2021-improved} + Diffusion Probabilistic Model & $>$54 (\textsuperscript{a}) 
& HiFi-GAN & 13.92  
& Whisper Medium (EN)
& 32\% \\

SU-E2S, Scheck.~\cite{Scheck2023SUE2S}
& HuBERT-Soft & N.A. 
& HiFi-GAN & 13.92 
& Whisper Medium (EN)
& 40\% \\

DiffMV-ETS~\cite{Scheck2025DiffMVETSDM}
& DiffMV-ETS & $\approx$ 66 \textsuperscript{(b)}
& BigVGAN & 112  
& Distil-Whisper-v3 
& N.A. \\

EMGVox-GAN~\cite{SUALIHEEN2025101754}
& as \cite{gaddy-klein-2021-improved} & as \cite{gaddy-klein-2021-improved}
& EMGVox-GAN & 12
& N.A. 
& 36\% \\

SWS, Lee~\cite{lee_speech_2025}
& as \cite{gaddy-klein-2021-improved} & 54 
& Sovits-SVC & N.A. 
& Whisper Medium (EN) 
& 38\% \\
\midrule

TinyMyo – FS 
& Transformer Encoder & 4.5M 
& HiFi-GAN & 13.92 
& Wav2Vec2 \cite{wav2wec} 
& 35.23$\pm$0.22\% \\

TinyMyo – FT 
& Transformer Encoder & 4.5M 
& HiFi-GAN & 13.92 
& Wav2Vec2 \cite{wav2wec} 
& 33.54$\pm$1.12\% \\

\midrule
TinyissimoMyo – FS
& Transformer Encoder & 2.6M
& HiFi-GAN & 13.92 
& Wav2Vec2 \cite{wav2wec} 
& 42.20$\pm$3.15\%\\

TinyissimoMyo – FT
& Transformer Encoder & 2.6M
& HiFi-GAN & 13.92 
& Wav2Vec2 \cite{wav2wec} 
& 40.94$\pm$3.43\% \\

\bottomrule
\end{tabular}

\vspace{2mm}
\parbox{0.95\linewidth}{
\footnotesize{
    \textsuperscript{(a)} = estimated. The authors employ the same Transduction model as in Gaddy \cite{gaddy-klein-2021-improved}, followed by a Diffusion Model (including residual blocks, attention and convolutional layers). 
    \textsuperscript{(b)} = estimated as sum of speaker encoder (6M), prior encoder (30M), and diffusion decoder (30M)} \\ 
    FS: model trained from scratch. FT: fine-tuning pretrained model
}
\end{table*}
Table \ref{tab:emg-synthesis} reports the results for the Speech Synthesis Task. 
We achieve a WER (FT) of $33.54\pm1.12 \%$, in line with previous works (SoA WER is 32\%, achieved by Ren et al. \cite{ren_speech_24}). 
Compared to previous solutions, our Transduction model offers a significant reduction in parameter size, with 62.5\% reduction compared to Scheck et al.  \cite{Scheck2023SUE2S} and 91.7\% reduction compared to Gaddy et al. \cite{gaddy-klein-2021-improved}. 
In this case, TinyMyo (FT) achieves 7\% performance improvement compared to the smaller version.

The reported number of parameters accounts only for Transduction model (encoder and model head). The number of parameters for the pre-trained Hi-Fi Gan vocoder is 13.92M (same as in \cite{gaddy-klein-2021-improved, ren_speech_24}). 
\\

\begin{table}[H]
  \centering
  \caption{Silent Speech Recognition Task - Comparison with SoA}
  \label{tab:emg-recognition}
  \begin{tabular}{l c c}
    \toprule
    \textbf{Method} & \textbf{Params} & \textbf{WER} \\
    \midrule
    Transformer, Gaddy \cite{gaddy_2022_phdthesis}      & 54 M        & 28.8\% \\
    \midrule
    TinyMyo - FT           & 4.5 M       & 33.95\%$\pm$0.97\% \\
    \midrule
    TinyissimoMyo - FT           &   2.6 M    & 43.07\%$\pm$1.03\% \\
    \bottomrule
  \end{tabular}
\vspace{1mm}

\parbox{0.8\linewidth}{
\footnotesize{
FT: fine-tuning pretrained model; \\
Benster et al \cite{benster_2024_mona} achieve 22.2\% and 12.2\% accuracy with the MONA and MONA LISA architectures, respectively. However, these results are obtained with multimodal pre-training (EMG and audio), without a specific breakdown on the contribution of individual modalities.}
}

\end{table}
Table \ref{tab:emg-recognition} reports the results for the Speech Recognition Task. Given the superior performance of the fine-tuned model compared to training from scratch in the Speech Production task, and the identical architecture adopted for the vocoder, in this case, we run the experiments only for the full fine-tuning setting.
We achieve a WER of $33.95\% \pm 0.97\%$, higher compared to 28.8\% reported of Gaddy et al. \cite{gaddy_2022_phdthesis}, however, with a 91.7\% reduction in parameter count.
As for the Speech Production Task, the WER achieved by the bigger model is significantly lower (9\% reduction) compared to the smaller variant.

The MONA LISA model \cite{benster_2024_mona} achieves a WER of 12.2\%. However, this performance relies on a multimodal training setup, where both EMG and audio data are jointly used. Moreover, MONA LISA is a large-scale language model, which poses practical limitations for real-time deployment and edge applications due to its substantial computational footprint and memory requirements.

\subsection{Model Deployment}
We deploy TinyMyo on the GAP9 microcontroller (MCU), as described in Section~\ref{sec:methods_deply}.
We achieve an inference time of $0.785~\text{seconds}$ (291.3 M cycles), with energy consumption of $44.90~\text{mJ}$ and average power-envelope of $57.18~\text{mW}$, with GAP9 operating at 370 MHz and using our custom deployment achieves a high compute utilization of 88.4\%; DMA load/store operations account for only $0.4\%$ of cycles, while idle time and overhead are limited to $11.2\%$. Crucially, we achieve a DMA/Compute overlap of $99.4\%$, effectively hiding data transfer latencies.

The model comprises 3.56M parameters, with the majority residing in the 8 transformer blocks. Each block contains a multi-head self-attention (MHSA) layer with 3 heads and a feed-forward network (FFN) with an expansion ratio of 4.

Table~\ref{tab:gap9_subop_tinymyo} summarize the runtime breakdown.
The overall utilization is high with limited DMA overhead. 
The total cycle count is dominated by the eight Transformer blocks.
The computational cost for one Transformer block (Table~\ref{tab:gap9_subop_tinymyo}) is dominated by the Multi-Head Attention (51.9\% of total cycles). In particular, the Q/K/V Projections, QK Attention Scores, and AV terms scale as $\mathcal{O}(N^2 d)$ with the sequence length $N=400$, making them the asymptotic computational bottleneck of the Transformer block. Crucially, the INT8 kernels reach strong throughput, with several attention and MLP sub-operations exceeding $9$--$10$ MACs/cycle (Table~\ref{tab:gap9_subop_tinymyo}, 
the MLP blocks in the model head account for 33.7\% of the total cycles). To overcome these runtime limitations, we also deploy the smaller TinyissimoMyo model variant (on 5-s and 1-s classification windows for the EPN-612 dataset), which has a lower sequence length (N=400 and N=80 respectively) and fewer transformer blocks (4). TinyissimoMyo achieves an inference latency of only \mbox{0.496 s} and \mbox{0.089 s} on 5-s and 1-s windows, respectively, thereby demonstrating low-latency compatible with real-time applications.

  \begin{table}[t]
  \vspace{-0.5cm}
  \caption{Sub-operation breakdown for one transformer block (TinyMyo, EMG input size 5 s)}
  \label{tab:gap9_subop_tinymyo}
  \resizebox{\columnwidth}{!}{
  \centering
  \begin{tabular}{lrrrr}
  \toprule
  \textbf{Operation} & \textbf{Cycles (M)} & \textbf{\%} & \textbf{MACs (M)} & \textbf{MACs/Cyc} \\
  \midrule
  Pre-Attention Norm & 2.2 & 6.2\% & 0.4 & 0.17 \\
  \textit{Multi-Head Attention} & \textit{18.6} & \textit{51.9\%} & \textit{120.4} & \textit{6.46} \\
  \quad Q/K/V Projections & 5.7 & 15.9\% & 44.2 & 7.76 \\
  \quad Permute/Reshape & 0.1 & $<$0.1\% & --- & --- \\
  \quad QK$^T$ Attention Scores & 3.4 & 9.4\% & 30.7 & 9.08 \\
  \quad Softmax Normalization & 4.5 & 12.6\% & --- & --- \\
  \quad Attention$\times$Values & 2.9 & 8.1\% & 30.7 & 10.61 \\
  \quad Output Projection & 1.9 & 5.3\% & 14.7 & 7.75 \\
  Pre-MLP Norm & 2.1 & 5.9\% & 0.4 & 0.18 \\
  MLP FC1 (192$\to$768) & 6.6 & 18.3\% & 59.0 & 8.97 \\
  GELU Activation & 0.8 & 2.3\% & 3.1 & 3.73 \\
  MLP FC2 (768$\to$192) & 5.5 & 15.4\% & 59.0 & 10.65 \\
  \midrule
  \textbf{Total} & \textbf{35.9} & \textbf{100\%} & \textbf{242.2} & \textbf{6.74} \\
  \bottomrule
  \end{tabular}
  }
  \end{table}
\section{Conclusion}\label{sec:conclusions}

We presented TinyMyo, the first compact EMG FM capable of broad generalization across diverse downstream tasks and real-time edge deployment on an ultra-low-power MCU.

TinyMyo is based on a Transformer-encoder architecture, with a size of 3.6 million parameters. The model is pre-trained using heterogeneous EMG data, collected from platforms with different numbers of channels and from diverse body locations. 

TinyMyo achieves SoA performance across multiple downstream tasks. For gesture recognition, we obtain SoA accuracy on the \textit{NinaPro DB5} ($87.98 \pm 0.96\,\%$), \textit{EPN-612} ($96.57 \pm 0.15\,\%$), and \textit{UCI-EMG} ($97.1 \pm 0.27\,\%$) datasets. Furthermore, on the \textit{Generic Neuromotor Interface -- discrete-gesture} dataset, TinyMyo achieves a CLER of $0.142\pm0.007$ with an approximate $44\,\%$ reduction in model size compared to SoA.

On the hand kinematic regression task, we achieve a MAE of $8.8 ^\circ \pm 0.23 ^\circ$ with a subject-agnostic model. 
On the speech production task, we obtain a WER of $33.54 \pm 1.12\,\%$, which is comparable to the current SoA, while reducing the model size by approximately $91\,\%$. Finally, we demonstrate that TinyMyo can also adapt to the speech recognition task, achieving a WER of $33.95\pm0.97\%$.

TinyMyo is also the first FM for EMG deployed on the edge, running in only $0.785~\text{s}$, with an energy of $44.91~\text{mJ}$ and an average power-envelope of $57.18~\text{mW}$. Furthermore, the smaller model variant (TinyissimoMyo) runs with an inference time as low as 0.089 s, thereby demonstrating compatibility with the low-latency requirements (sub-100 ms) of real-time applications such as prosthesis control (the accuracy-latency trade-off is discussed below). Additionally, the model supports overlapping causal sliding windows, allowing the stride to be chosen independently of the input-window length. TinyissimoMyo requires approximately 89~ms for an inference pass with a 1s input window. Therefore, predictions can be generated at intervals substantially shorter than the input-window duration, as long as the inference time remains below the selected stride.\footnote{For example, on the Ninapro DB5 dataset, using 1~s windows with 80\% overlap (i.e., 200~ms stride), TinyissimoMyo achieves an accuracy of $89.97 \pm 0.65\%$. Since the inference time (89~ms) is considerably shorter than the stride (200~ms), each prediction can be completed before the next input window becomes available.} This enables the use of longer input windows to preserve temporal context and classification performance without constraining the prediction rate to one output per input-window duration.

Future work will focus on the following limitations:
\begin{itemize}
    \item The datasets used for pre-training consist primarily of EMG recordings from the forearm and wrist during gesture-related tasks. Data from additional body locations (e.g., lower-limb muscle groups) should be included to improve anatomical generalization.
    \item Both the pre-training and downstream datasets were collected in controlled laboratory environments. Additional research is needed to evaluate the robustness of the proposed FM under realistic conditions, such as data corrupted by motion artifacts or environmental noise.
    \item Further studies are required to assess the model’s generalization capabilities in zero-shot settings and to quantify the amount of labeled data needed to achieve strong performance on novel downstream tasks.
    \item Despite demonstrating the first deployment of an FM for EMG on an ultra-low power MCU, further improvements are still possible. On the EPN-612 deployment benchmark, the inference time was reduced from 0.785 s (TinyMyo, 5-s window) to 0.496 s (TinyissimoMyo, 5-s window) with a small (3\%) accuracy drop, and further to 0.089 s on 1-s windows at the cost of a larger accuracy drop (reaching an accuracy of $\approx$72\%). This highlights the accuracy-latency trade-off within the proposed model family and input window size.
    Future work will therefore focus on reducing the computational cost of inference while preserving high accuracy through targeted architectural optimization. This could be achieved by doing an ablation study on reducing the sequence length, by adopting a cheaper attention mechanism~\cite{dimofte_2025_cerebro},  more aggressive quantization techniques (e.g., sub 8-bit \cite{dong_2023_packqvit}, and evaluating state-space models \cite{gu_2024_mamba,tegon_2025_femba}, whose computational cost scales linearly with the context length.
    Also, distributing the computation over multiple GAP9 nodes or deploying it on more powerful platforms~\cite{prasad_2024_siracusa} would yield substantial speed-ups.

\end{itemize}

By open-sourcing{\footnote{\url{https://github.com/pulp-bio/BioFoundation}} our pre-trained models and the architectures used for downstream evaluations, we aim to provide a versatile resource that can accelerate future research and serve as a foundation for the broader EMG community. Our intention is that other researchers will extend this backbone with additional datasets, improved training strategies, and architectural variations.

\section*{Acknowledgment}
We thank Chen Yanlong (ETH Z{\"u}rich) for technical support and Lorenzo Lamberti  (ETH Z{\"u}rich) for fruitful discussions. 
\section*{References}
\bibliographystyle{IEEEtran}
\bibliography{biblio}
\end{document}